\documentclass[a4paper,twoside]{article}

\usepackage{epsfig}
\usepackage{subcaption}
\usepackage{calc}
\usepackage{amssymb}
\usepackage{amstext}
\usepackage{amsmath}
\usepackage{amsthm}
\usepackage{multicol}
\usepackage{pslatex}

\usepackage{color,soul}

\usepackage{multirow}
\usepackage{diagbox}

\usepackage{algorithm}
\usepackage{algpseudocode}

\usepackage{hyperref}

\usepackage{natbib}
\let\bibhang\relax
\usepackage{apalike}

\usepackage{SCITEPRESS}     

\begin{document}

\title{RTSDF: Real-time Signed Distance Fields for Soft Shadow Approximation in Games}

\author{\authorname{Yu Wei Tan\orcidAuthor{0000-0002-7972-2828}, Nicholas Chua, Clarence Koh and Anand Bhojan\orcidAuthor{0000-0001-8105-1739}}
\affiliation{School of Computing, National University of Singapore, Singapore}
\email{\{yuwei, nicholaschuayunzhi, clarence.koh\}@u.nus.edu, banand@comp.nus.edu.sg}
}

\keywords{Real-time, Signed Distance Field, Jump Flooding, Ray Tracing, Soft Shadow, Rendering, Games.}

\abstract{Signed distance fields (SDFs) are a form of surface representation widely used in computer graphics, having applications in rendering, collision detection and modelling. In interactive media such as games, high-resolution SDFs are commonly produced offline and subsequently loaded into the application, representing rigid meshes only. This work develops a novel technique that combines jump flooding and ray tracing to generate approximate SDFs in real-time. Our approach can produce relatively accurate scene representation for rendering soft shadows while maintaining interactive frame rates. We extend our previous work with details on the design and implementation as well as visual quality and performance evaluation of the technique.}

\onecolumn \maketitle \normalsize \setcounter{footnote}{0} \vfill

\section{\uppercase{Introduction}}

Signed distance fields (SDFs) are scalar fields that store the shortest distance between a point in space to a model. Their sign indicates if that point is inside or outside the bounds of said model. In interactive media, models are most commonly represented by triangle meshes. SDFs are typically generated offline through ray tracing and scan conversion etc., limiting their use to rigid meshes. While existing real-time GPU-based methods can update SDFs per frame, they are unable to handle high resolutions efficiently.

We present a novel SDF method that integrates jump flooding and ray tracing to generate an approximate real-time SDF (RTSDF) of reasonably high quality for a fixed small scene. Additionally, we evaluate the technique by applying it to raymarched soft shadow approximation, offering trade-offs between speed and quality for real-time application requirements. This paper extends our previous work \citep{Tan:2020:RGS} with a detailed analysis of the design, implementation and evaluation of the technique.
 \section{\uppercase{Design}}

\begin{figure}[!h]
    \centering
    \subcaptionbox{Jump flooded SDF}{
        \includegraphics[width=0.47\linewidth]{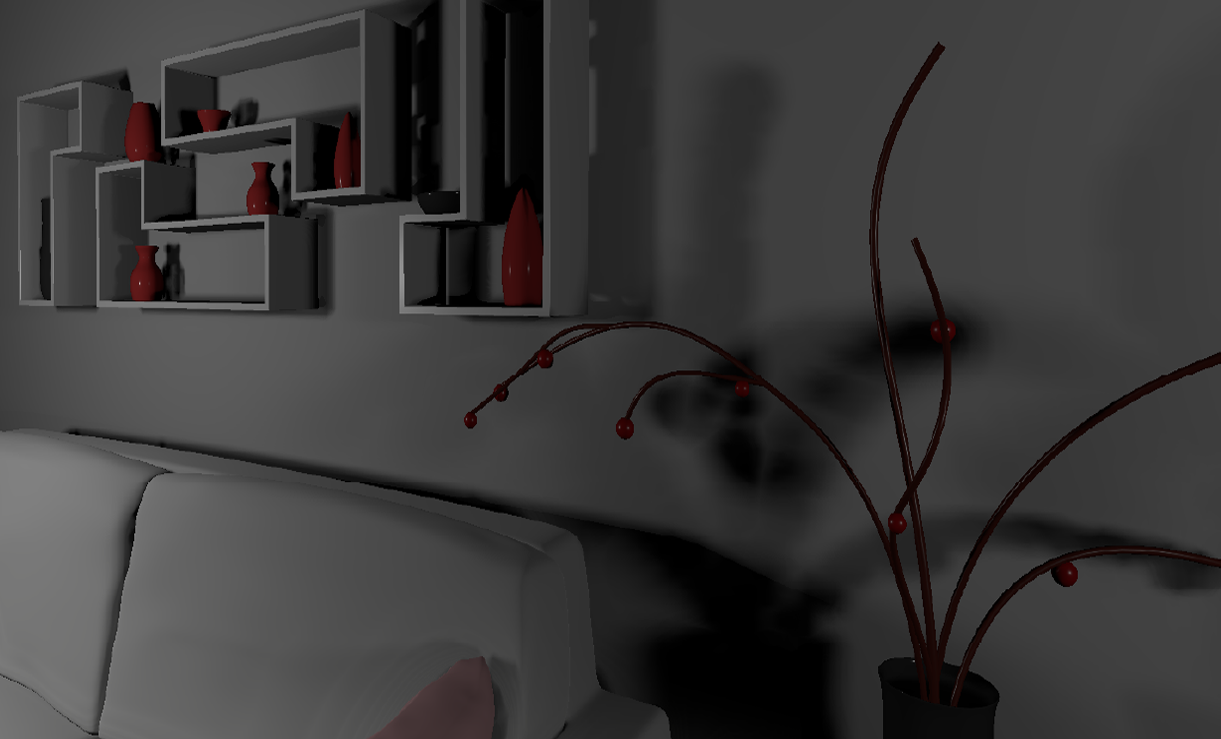}
    }
    \subcaptionbox{Ray-traced SDF}{
        \includegraphics[width=0.47\linewidth]{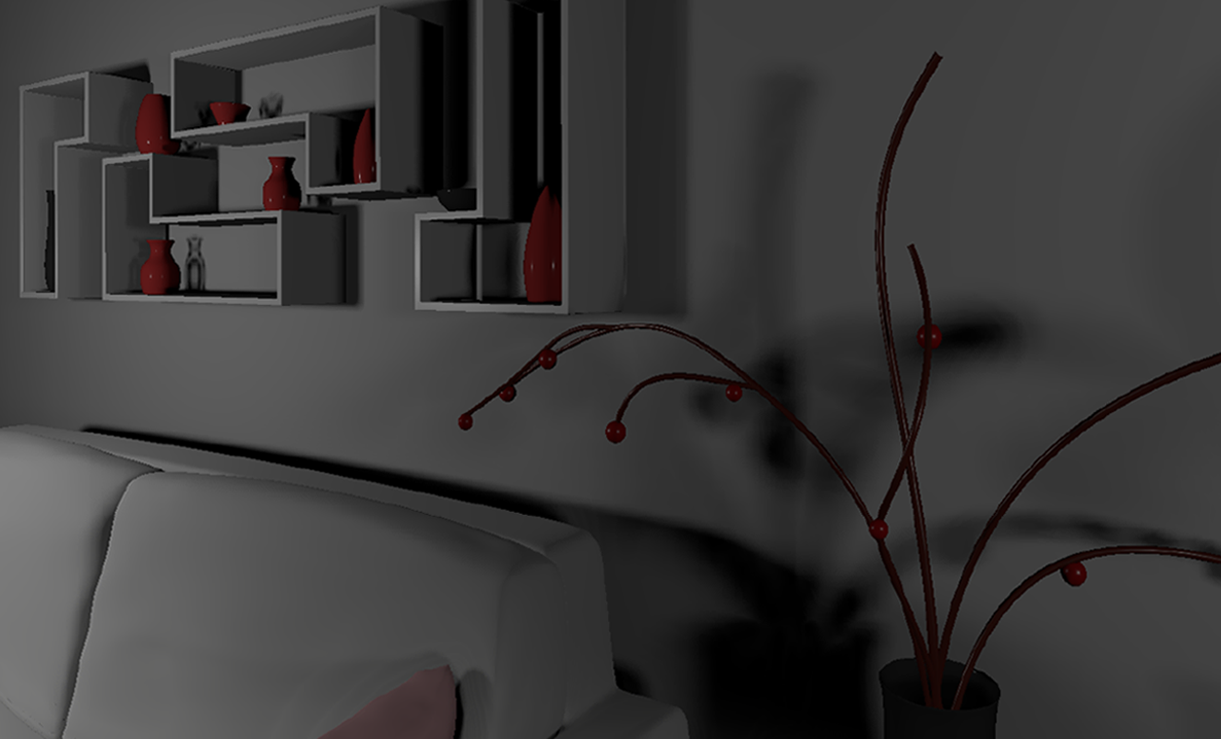}
    }
    \par\smallskip
    \caption{Soft shadows from SDF raymarching.}
    \label{fig:soft-shadows}
\end{figure}

As shown in \autoref{fig:soft-shadows}, jump flooding produces a fast approximation of the SDF which allows for real-time calculation. Conversely, ray tracing gives a more accurate scene representation as it queries the hardware-generated triangle mesh and slowly converges. Hence, we propose a real-time SDF that combines the speed of jump flooding with the precision of ray tracing. We first perform an initial jump flooding which creates an SDF of the voxelized scene. Next, we use the voxelized SDF as a mask to decide where to attempt ray tracing on the triangle mesh for more accurate scene representation. Naturally, we choose locations closer to surfaces to ray trace and fall back on the distances generated by jump flooding in empty regions. Our technique is built on NVIDIA’s Falcor library \citep{NVIDIA:2017:FRF} which provides an abstraction over the graphics API for our implementation.

\subsection{Ray Tracing}

The basic implementation of an SDF is a uniform grid where the discrete points in the scalar field are stored as a 3D texture \citep{Wright:2015:ARR}. Uniform grids are easy to implement and allow us to perform hardware interpolation of neighbouring points, making sampling efficient. A brute force approach to generate uniform SDF is ray sampling \citep{Wright:2015:ARR}. 


For each discrete point in the distance field, we shoot rays in random directions to query the distance to the closest mesh. To calculate the direction, we randomly generate a point on the surface of a sphere with a uniform distribution \citep{Weisstein:2019:SPP}. The minimum distance traced for each point is stored in the red channel of a $400 \times 200 \times 400$ 3D texture. Rays traced in future frames will overwrite the value if the newer value is smaller. To determine the sign, we first check if it is a front or back face hit via the dot product of the ray direction and normal of the primitive intersected. We then accumulate the number of front and back face hits in the green and blue channels of the texture. Finally, we set the sign to negative if the majority of hits are back face hits \citep{Wright:2015:ARR}. 








\subsection{Jump Flooding}

The jump flooding algorithm (JFA) \citep{Rong:2006:JFG} which can be run on parallel on the GPU allows us to generate an approximate SDF in real-time. We first initialize a 3D SDF texture where each texel represents a 3D point or a grid point. JFA gives us information about the closest seed at any point in space. Setting every point on each triangle in the mesh as a seed, we can obtain the closest distance to a surface for our distance field. To determine if a grid point contains a triangle efficiently, we voxelize the scene with voxel resolution equal to our final distance field. After voxelization, we simply check if a grid point contains a model voxel to determine if it is a seed.




We now have a 3D texture containing grid points that are either empty or are seeds. Without loss of generality, we assume that the dimensions of the 3D texture are equal (i.e. 3D cube) and that its length $n$ is a power of 2. For each grid point, we query a constant number of neighbouring grid points a predefined offset away. For each query, if the queried grid point is a seed or contains seed information, we check if that seed is closer to its currently stored seed and update its seed information if so. We start with an offset of length $\frac{n}{2}$ and halve it for each subsequent iteration until it reaches 1 for our final iteration. 



With current GPUs that can write to 3D textures, we adapt the 2D JFA algorithm \citep{Rong:2006:JFG} for 3D space. During a single iteration, we run the querying in parallel, allowing us to use the GPU to accelerate the calculations. However, the algorithm produces an unsigned distance field. To determine the sign, we subtract a small $\beta$ from the distance field, causing surface points to be of negative value and effectively thickening the surface. The surfaces generated are also hollow as they contain positive values in their interior. With jump flooding, we can generate an approximate SDF in real-time for a decently large resolution of $256^3$ at 30ms and $128^3$ at 2.34ms.



\subsection{Ray Mask}

\begin{figure}[!h]
    \centering
    \subcaptionbox{\emph{Coarse} SDF}{
        \includegraphics[width=0.47\linewidth]{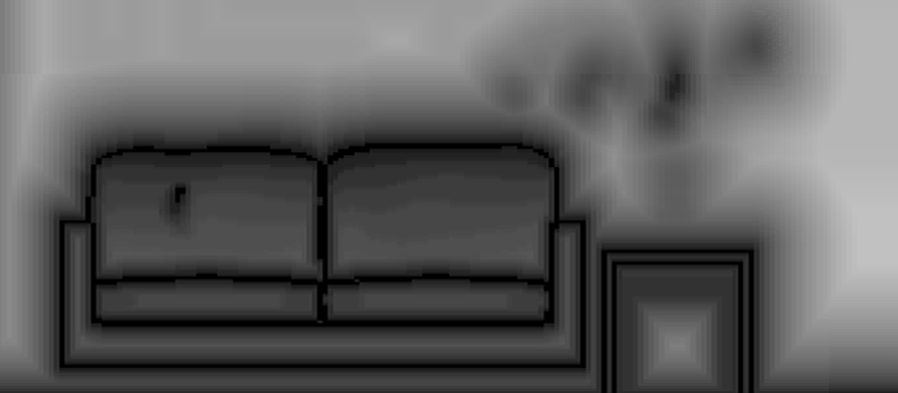}
    }
    \subcaptionbox{\emph{Fine} SDF}{
        \includegraphics[width=0.47\linewidth]{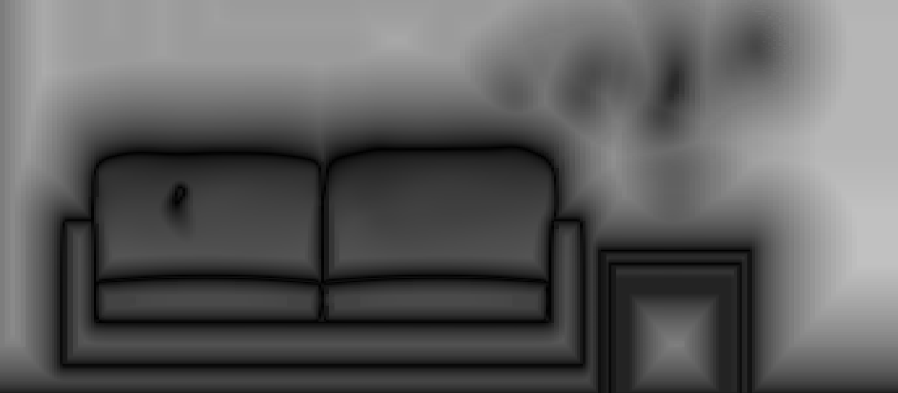}
    }
    \par\smallskip
    \caption{Slice of SDF.}
    \label{fig:sdf-slice}
\end{figure}

We obtain a rough approximation of the SDF or \emph{coarse} SDF via jump flooding to locate regions in the scene to apply ray tracing. In raymarching, regions far from the surface act as a way to accelerate the process but closer regions require a more accurate surface representation. Hence, we can detect regions closer to surfaces based on a distance $d$ from the \emph{coarse} SDF and only ray trace in these regions at a higher resolution for better surface representation. Essentially, the \emph{coarse} SDF is a ray mask that determines which areas in the SDF should be ray-traced to generate a \emph{fine} SDF as shown in \autoref{fig:sdf-slice}. $d$ can be used to trade-off performance for accuracy where a larger $d$ results in more rays traced as texels further from surfaces would be within $d$ distance from a surface point.

Unlike adaptively sampled fields which increase the resolution at regions with finer details, we limit the number of levels of detail to two as generation of hierarchical SDF is difficult to parallelize \citep{Liu:2014:EAS}. Additionally, real-time traversal of the SDF may require multiple texture lookups to sample until the leaf node despite saving space with a sparse voxel texture \citep{Aaltonen:2018:AGT}. 

\begin{figure}[!h]
    \centering
    \subcaptionbox{\emph{Coarse} SDF}{
        \includegraphics[width=0.47\linewidth]{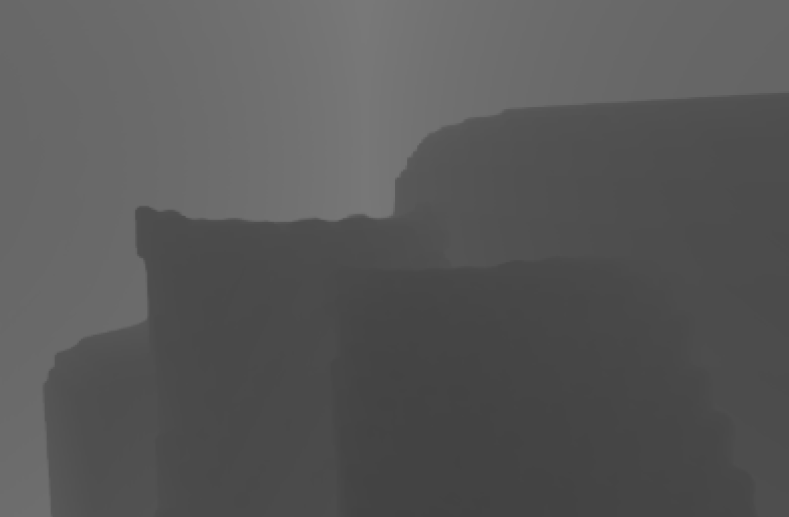}
    }
    \subcaptionbox{\emph{Fine} SDF}{
        \includegraphics[width=0.47\linewidth]{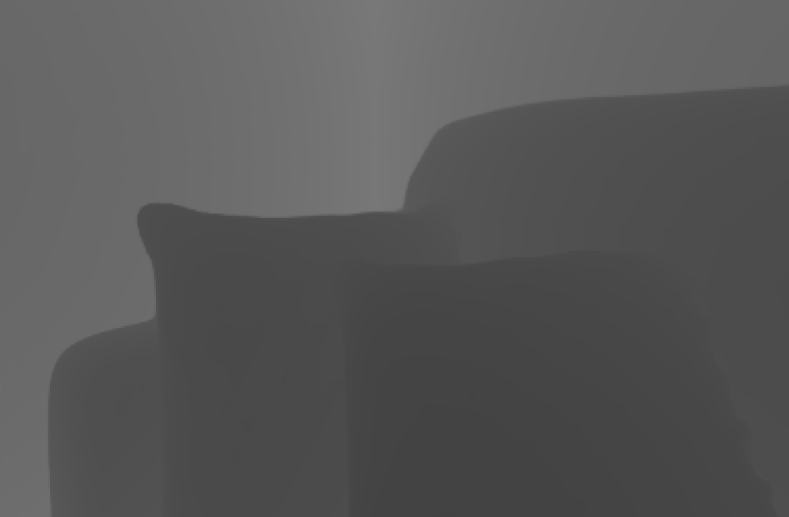}
    }
    \par\smallskip
    \caption{Precision.}
    \label{fig:couch}
\end{figure}
\begin{figure}[!h]
    \centering
    \subcaptionbox{\emph{Coarse} SDF}{
        \includegraphics[width=0.47\linewidth]{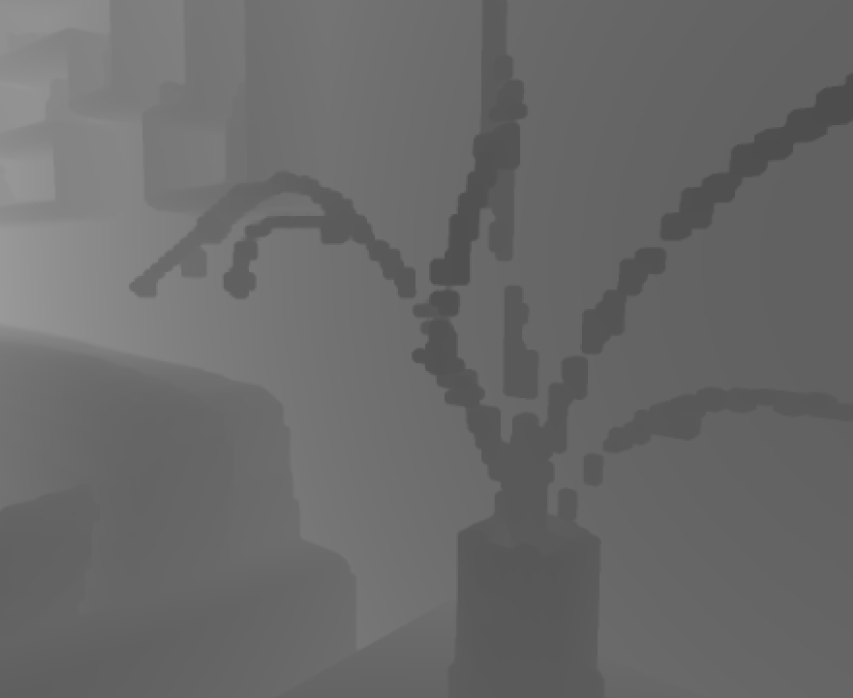}
    }
    \subcaptionbox{\emph{Fine} SDF}{
        \includegraphics[width=0.47\linewidth]{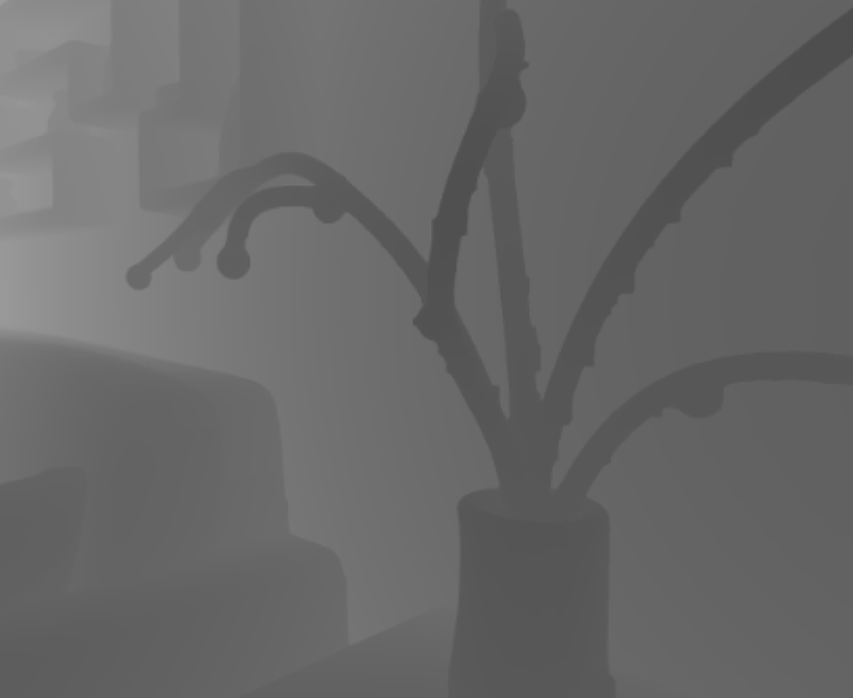}
    }
    \par\smallskip
    \caption{Holes.}
    \label{fig:holes-thin-surface}
\end{figure}

With the combination of the techniques, we turn a blocky representation of the scene into a more refined triangle mesh representation without ray tracing every texel in the SDF as shown in \autoref{fig:couch}. Additionally, we can resolve issues identified in the jump flooding of voxelization of thin surfaces. As seen in \autoref{fig:holes-thin-surface}, while the \emph{coarse} SDF gets a disconnected representation of the plant, we fill in the holes through refinement with the ray trace pass.

\subsection{Raymarching}

While SDFs are surface representations, they only provide proximity information without a direction. To perform ray casting on an SDF, we can find the intersection with raymarching. The SDF value shows the safest possible distance we can move along the ray without missing any potential intersections. Hence, given a ray origin and direction, we can query the SDF at safe points along the ray and terminate if the distance field returns a value less than or equal to 0, signalling that we have reached a surface point. 

This raymarching algorithm, also known as sphere tracing \citep{Hart:1996:STA}, does not account for no intersection and could potentially iterate indefinitely in practice. A simple solution would be to terminate if the raymarched distance exceeds the bounds of the queried scene. More commonly, we can set a maximum number of iterations to perform and return no intersection if the limit is reached, ensuring consistent execution time for interactive rendering. Additionally, as the raymarch count approaches $\infty$, we are marching closer to a surface but will never reach it, potentially making the algorithm run indefinitely. Hence, we could terminate the algorithm if the closest distance to a surface is lower than some $\epsilon$. 





SDFs are used to calculate dynamic occlusion in rendering like for area light shadow approximation as in \citet{Wright:2015:ARR}. Considering direct illumination of the area light source alone, we can reform the rendering equation as explained in \citet{Dutre:2004:SAM}. \citet{Wright:2015:ARR} uses raymarching of SDFs to approximate the visibility term for the entire area light source.

With ray tracing, we can only get the intersection point with the surface. However, we can also know the closest it got to intersecting an object with raymarching. To get this information with ray tracing, we need to shoot multiple rays within a cone which is more accurate but too inefficient for real-time.

\subsection{Correcting Artifacts}






\subsubsection{Ghosting}

\begin{figure}[!h]
    \centering
    \includegraphics[width=\linewidth]{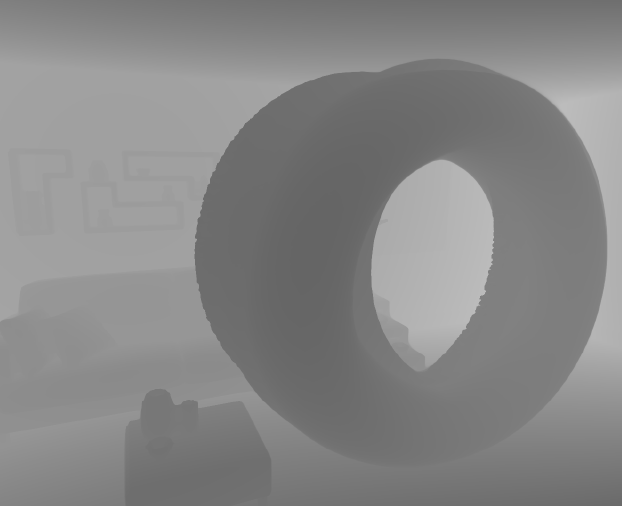}
    \caption{Ghosting of circular-moving teapot.}
    \label{fig:sdf-ghosting}
\end{figure}

While the \emph{fine} SDF generated is comparable to full ray sampling, it uses temporal accumulation to converge to the final distance field. Consequently, when a moving object is introduced in the scene, its surface ghosts as shown in \autoref{fig:sdf-ghosting} because the area it leaves is not invalidated as the closest distance.

\begin{figure}[!h]
    \centering
    \subcaptionbox{$128^3$ \emph{coarse} SDF input}{
        \includegraphics[width=0.47\linewidth]{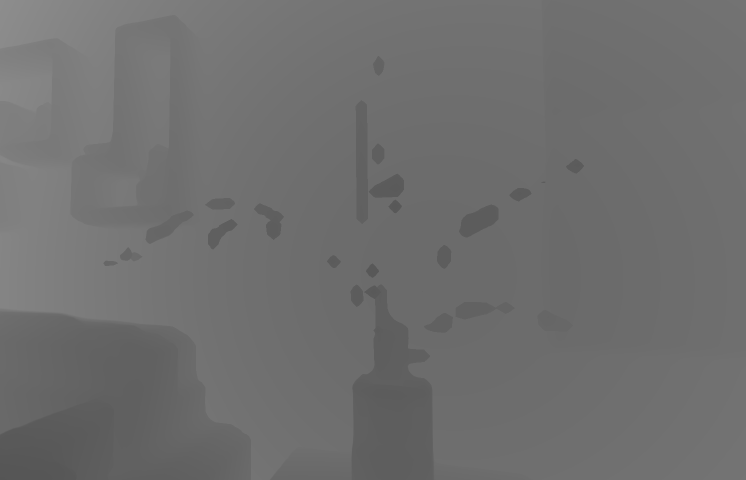}
    }
    \subcaptionbox{$x$ = 1\label{fig:jump-flood-1ray}}{
        \includegraphics[width=0.47\linewidth]{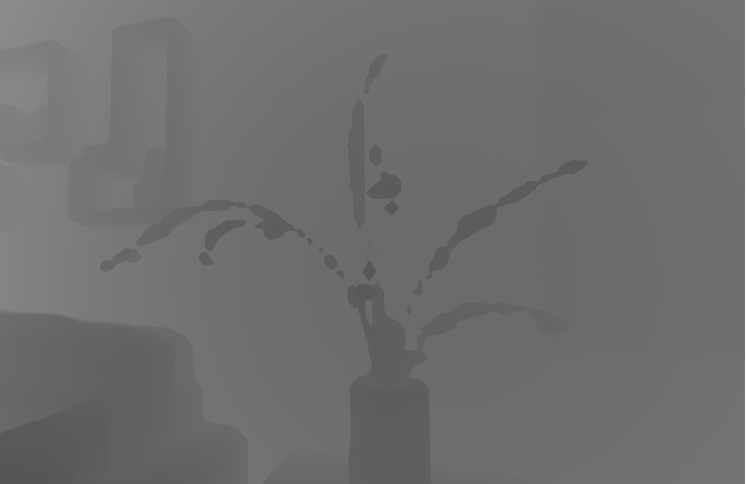}
    }
    \par\smallskip
    \subcaptionbox{$x$ = 5\label{fig:jump-flood-5rays}}{
        \includegraphics[width=0.47\linewidth]{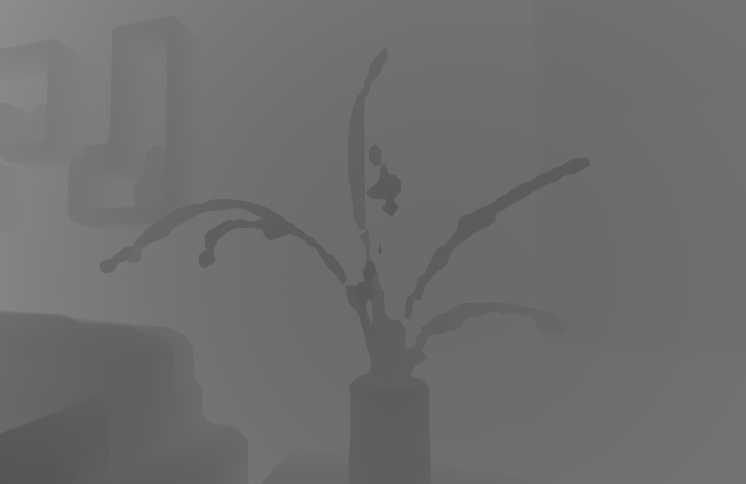}
    }
    \subcaptionbox{$x$ = 10\label{fig:jump-flood-10rays}}{
        \includegraphics[width=0.47\linewidth]{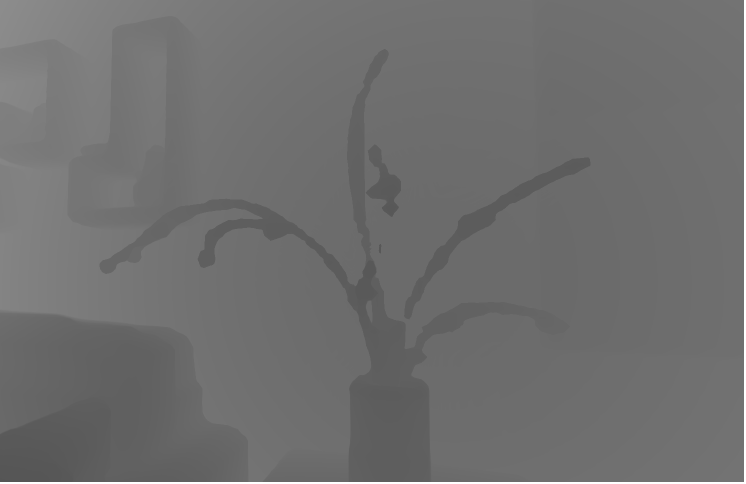}
    }
    \par\smallskip
    \caption{\emph{Fine} SDF result.}
    \label{fig:rpp}
\end{figure}

As such, each texel in the SDF must be recalculated every frame. However, to get stable results, the number of rays $x$ shot per frame must be substantially high. Otherwise, it will result in noisy SDF as shown in \autoref{fig:rpp} due to random sampling - in each frame, there is a chance we might not hit the closest surface. Conversely, the value of $x$ is restricted by the computational cost of ray tracing. As a compromise, we spread our effective rays shot over multiple frames with temporal accumulation of the SDF but apply a decay factor to the distance field in the previous frame to minimize ghosting as shown in \autoref{equ:decay}.

\begin{equation}
  f_{t} = 
  \begin{cases} 
    \min(\alpha \cdot f_{t - 1} + (1 - \alpha) \cdot c_{t}, r_{t}), & \text{for } c_{t} \leq d \\
    c_{t}, & \text{for } c_{t} > d
  \end{cases}
  \label{equ:decay}
\end{equation}

Let $f_t$, $c_t$ and $r_t$ be the \emph{fine} SDF, \emph{coarse} SDF and the shortest distance generated from ray sampling respectively at frame $t$. $\alpha$ refers to the decay factor for the previous frame. Note that we eliminate any form of ghosting for $c_t > d$ as we use $c_t$, the jump flooded SDF that is newly calculated every frame. To perform jump flooding every frame, the \emph{coarse} SDF is set to $128^3$ while the \emph{fine} SDF is at a resolution of $256^3$. Our testing found an $\alpha$ of 0.95 and $d$ of 0.1 provides a good qualitative result which minimizes ghosting of moving surfaces. 

\subsubsection{Banding}

\begin{figure}[!h]
    \centering
    \subcaptionbox{Before triangulation}{
        \includegraphics[width=0.3\linewidth]{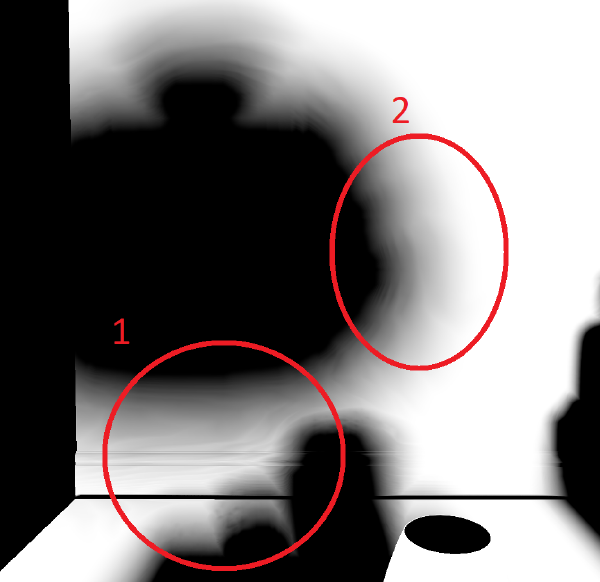}
    }
    \subcaptionbox{After triangulation}{
        \includegraphics[width=0.3\linewidth]{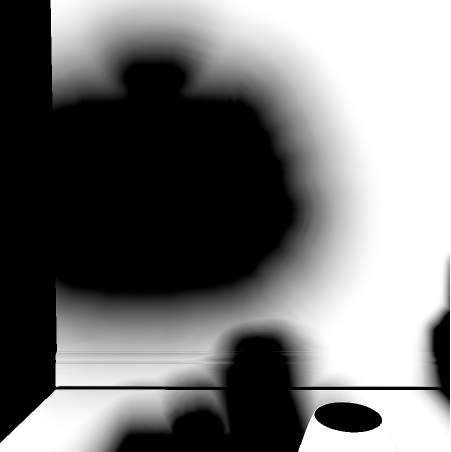}
    }
    \subcaptionbox{After triangulation and max. step size}{
        \includegraphics[width=0.3\linewidth]{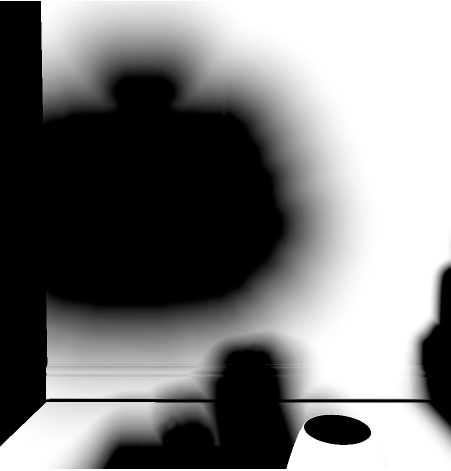}
    }
    \par\smallskip
    \caption{Banding from low resolution SDF.}
    \label{fig:teapot-banding}
\end{figure}

Banding artifacts come from the low resolution of the SDF for real-time optimization. They appear along the penumbra of shadows, causing alternating regions of high and low occlusion as shown in \autoref{fig:teapot-banding}.

\begin{figure}[!h]
    \centering
    \includegraphics[width=0.9\linewidth]{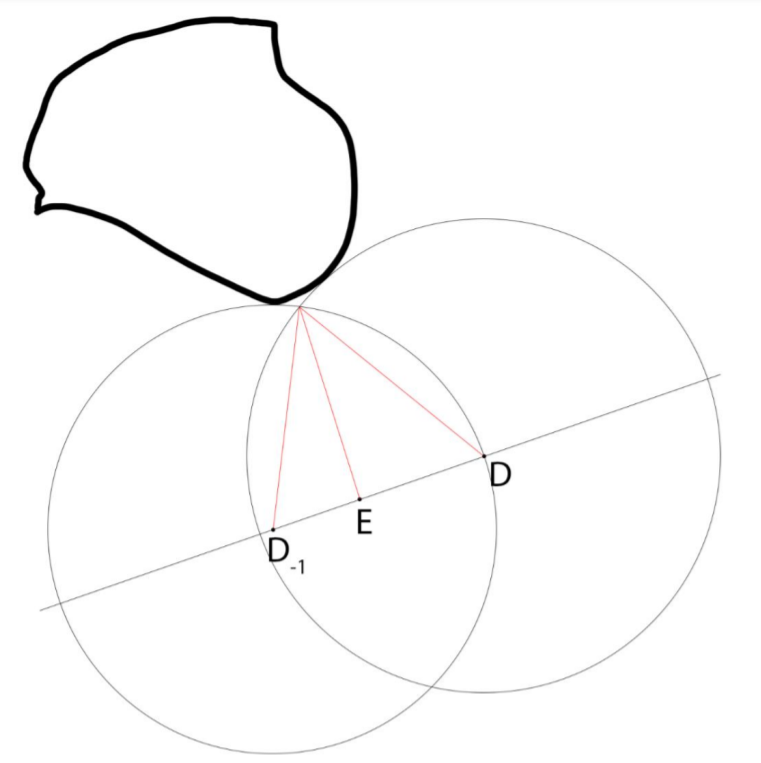}
    \caption{Approximation of occlusion with triangulation. Source: \citet{Aaltonen:2018:AGT}.}
    \label{fig:approx-triangulation}
\end{figure}

We reduce some of the banding by approximating the occlusion according to \citet{Aaltonen:2018:AGT}. From \autoref{fig:approx-triangulation}, given two raymarch samples $D$ and $D_{-1}$, we can triangulate to calculate an approximation of the SDF at $E$ which would have contributed to a higher occlusion than both samples. 

To ensure that neighbouring pixels have similar raymarch samples, we also restrict the maximum step size of our raymarching to 0.05 to remove the banding. However, a consequence would be an increase in texture samples (128 in this case) required to calculate occlusion so we are looking into finding a better compromise. Nonetheless, we achieve a smooth penumbra for now, as expected of a soft shadow.

\begin{figure}[!h]
    \centering
    \subcaptionbox{Before TAA}{
        \includegraphics[width=0.47\linewidth]{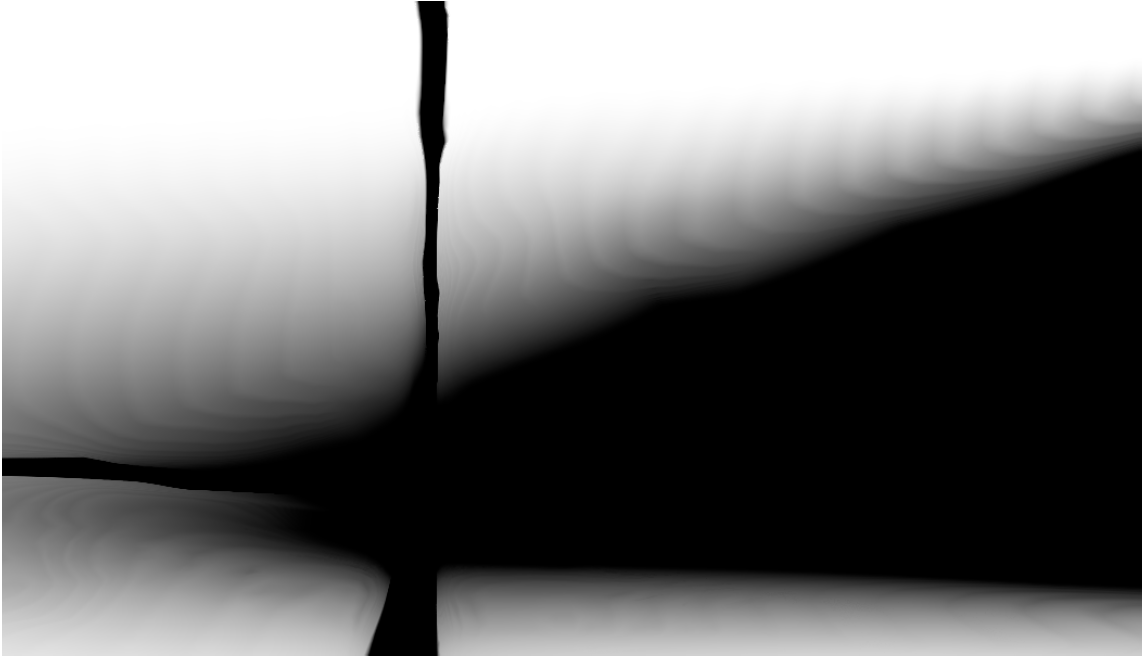}
    }
    \subcaptionbox{After TAA}{
        \includegraphics[width=0.47\linewidth]{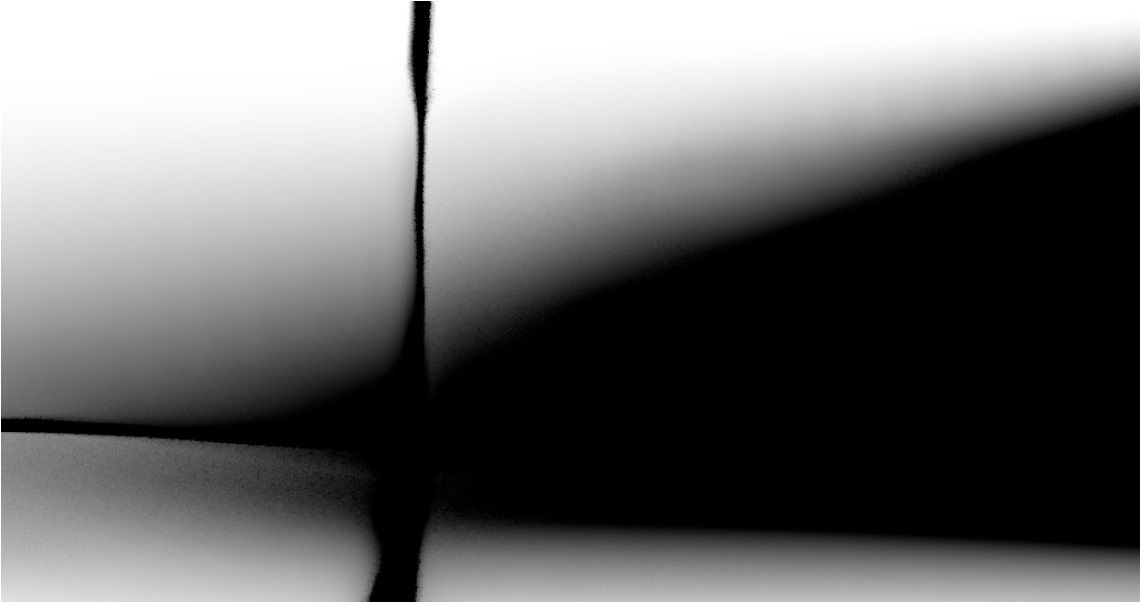}
    }
    \par\smallskip
    \caption{Banding from fixed step size.}
    \label{fig:sofa-banding}
\end{figure}

Lastly, due to the fixed step size, there is also an obvious pattern when shadows are parallel to the light direction. We remove this sampling artifact by jittering the offset of the ray so that we move our ray sample slightly along the light direction. We then apply a temporal anti-aliasing (TAA) \citep{Karis:2014:HQT} pass to remove any noise in the final image for a smooth gradient as shown in \autoref{fig:sofa-banding}. 

\subsubsection{Holes}
\label{sec:holes}

\begin{figure}[!h]
    \centering
    \subcaptionbox{Low resolution}{
        \includegraphics[width=0.47\linewidth]{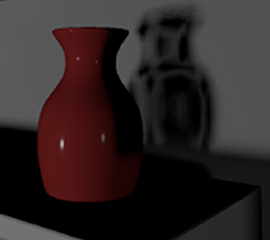}
    }
    \subcaptionbox{High resolution}{
        \includegraphics[width=0.47\linewidth]{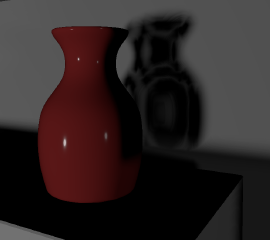}
    }
    \par\smallskip
    \caption{Holes in SDF.}
    \label{fig:sdf-holes}
\end{figure}

Holes in our SDFs are observable from the shadow of objects with thin surfaces like in \autoref{fig:sdf-holes} due to the low resolution of the SDF. Increasing the resolution to 500 $\times$ 500 $\times$ 500 only reduces the size of the holes but does not eliminate them, and results in a poor rendering performance of 70ms. Consequently, we add a bias of 0.01 to our SDF to thicken surfaces. With this change, we compromise the accuracy of the surface representation for cleaner shadows. 

\begin{figure}[!h]
    \centering
    \includegraphics[width=\linewidth]{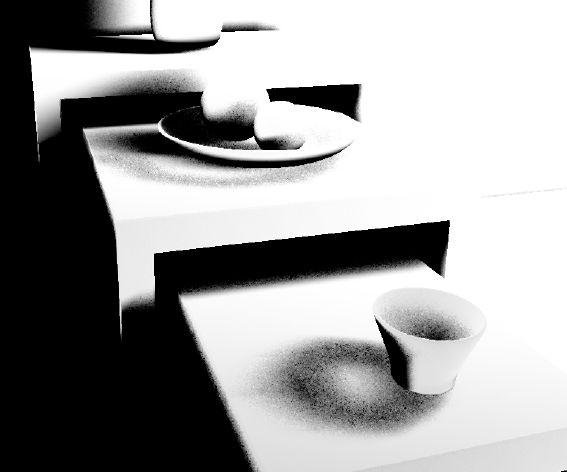}
    \caption{Holes in umbra.}
    \label{fig:umbra-holes}
\end{figure}

Another artifact comes from the $\epsilon$ of our raymarching technique. $\epsilon$ must be minimally one voxel size to avoid self-occlusion. However, the ray steps through thin surfaces entirely due to the large $\epsilon$ arising from our low-resolution distance field, appearing as holes in the shadow umbra as shown in \autoref{fig:umbra-holes}. A solution is to combine our soft shadow technique with a classic hard shadow approach like Cascaded Shadow Maps (CSM) \citep{Engel:2006:CSM} which can achieve the shadow umbra.
\section{\uppercase{Results}}

We test our RTSDF technique by generating raymarched soft shadows and comparing them with the shadows produced by shadow mapping and ground truth distributed ray tracing \citep{Cook:1984:DRT}. Our shadow map implementation makes use of the CSM and Exponential Variance Shadow Maps (EVSM) \citep{Lauritzen:2008:LVS} filtering sample provided by Falcor. We perform our evaluation on \textsc{The Modern Living Room} \citep{Wig42:2014:MLR} with dynamic objects and TAA.



\subsection{Performance}

The measurements here are taken with the Falcor profiling tool on an Intel Core i7-8700K CPU at 16GB RAM with an NVIDIA GeForce RTX 2080 Ti GPU.

\subsubsection{Comparison with Shadow Mapping}

We evaluate the performance of RTSDF with one directional light as shown in \autoref{tab:hybridsdf}. Without extensive optimizations on the Falcor API and Direct3D level, we are already achieving relatively interactive frame rates and pass durations. In comparison, shadow mapping has a frame rate of 241 fps, but it is expected that our method is slower because of our additional SDF generation and raymarching processes. We are also using additional abstractions and wrappers provided by \citet{Wyman:2018:IDR} for ease of implementation. 


\begin{table}[!h]
\begin{center}
    \caption{RTSDF pass durations (ms) and frame rate.}
    \label{tab:hybridsdf}
\begin{tabular}{|l|c|c|c|c|}
    \hline
    Passes                                   & Processor & Duration \\ \hline
                                             & CPU       & 0.29     \\ \cline{2-3}
    \multirow{-2}{*}{G-Buffer}               & GPU       & 1.05     \\ \hline
                                             & CPU       & 0.63     \\ \cline{2-3}
    \multirow{-2}{*}{Voxelization (V)}       & GPU       & 0.93     \\ \hline
                                             & CPU       & 0.06     \\ \cline{2-3}
    \multirow{-2}{*}{Jump Flood (JF)}        & GPU       & 2.09     \\ \hline
                                             & CPU       & 1.05     \\ \cline{2-3}
    \multirow{-2}{*}{Ray Trace (RT)}         & GPU       & 4.60     \\ \hline
                                             & CPU       & 0.11     \\ \cline{2-3}
    \multirow{-2}{*}{Deferred Lighting (DL)} & GPU       & 1.28     \\ \hline
                                             & CPU       & 7.42     \\ \cline{2-3}
    \multirow{-2}{*}{Others}                 & GPU       & 0.27     \\ \hline
                                             & CPU       & 9.56     \\ \cline{2-3}
    \multirow{-2}{*}{Total Duration}         & GPU       & 10.22    \\ \hline
    \multicolumn{2}{|l|}{Frame Rate}         & \textbf{97}          \\ \hline
\end{tabular}
\end{center}
\end{table}


\subsubsection{SDF Resolution}

We compare the increasing size of our SDF and record the GPU timings for the SDF generation passes and final deferred lighting computation on three lights at different resolutions specified in \autoref{tab:resolutions}. 


\begin{table}[!h]
\begin{center}
    \caption{Resolution for different SDF sizes.}
    \label{tab:resolutions}
\begin{tabular}{|l|c|c|}
    \hline
    Resolution & \emph{Coarse} SDF & \emph{Fine} SDF \\ \hline
    Small (S)  & $64^3$            & $128^3$         \\ \hline
    Medium (M) & $128^3$           & $256^3$         \\ \hline
    Large (L)  & $256^3$           & $512^3$         \\ \hline
\end{tabular}
\end{center}
\end{table}

As seen in \autoref{tab:resolution}, the duration of jump flooding becomes much higher with increasing SDF resolution, as there are more dispatch calls when executing the compute shader as well as reduced GPU cache locality when reading and writing to a much larger 3D texture. Jump flooding for $256^3$ \emph{coarse} SDF costs 21.89ms which results in a frame rate of lower than 60 frames per second. Consequently, we note that with current optimizations applied to jump flooding, we are limited to $128^3$ resolution. Nonetheless, with a lower \emph{fine} SDF resolution of $256^3$, we can afford to shoot more rays per texel which improves the stability of the distance field generated.

\begin{table}[!h]
    \caption{GPU timings of passes (ms).}
    \label{tab:resolution}
\begin{center}
\begin{tabular}{|c|c|c|c|c|c|}
    \hline
    Size                 & $x$ & 0     & 1     & 5     & 10    \\ \hline
                         & S   & 0.49  & 0.49  & 0.49  & 0.49  \\ \cline{2-6}
                         & M   & 0.49  & 0.49  & 0.49  & 0.49  \\ \cline{2-6}
    \multirow{-2}{*}{V}  & L   & 1.12  & 1.12  & 1.12  & 1.12  \\ \hline
                         & S   & 0.28  & 0.28  & 0.28  & 0.28  \\ \cline{2-6}
                         & M   & 2.37  & 2.37  & 2.37  & 2.37  \\ \cline{2-6}
    \multirow{-2}{*}{JF} & L   & 21.89 & 21.89 & 21.89 & 21.89 \\ \hline
                         & S   & 0.14  & 0.16  & 0.71  & 1.41  \\ \cline{2-6}
                         & M   & 0.55  & 1.41  & 4.51  & 8.36  \\ \cline{2-6}
    \multirow{-2}{*}{RT} & L   & 3.98  & 9.12  & 26.56 & 48.32 \\ \hline
                         & S   & 3.17  & 3.17  & 3.17  & 3.17  \\ \cline{2-6}
                         & M   & 3.17  & 3.17  & 3.17  & 3.17  \\ \cline{2-6}
    \multirow{-2}{*}{DL} & L   & 7.32  & 7.32  & 7.32  & 7.32  \\ \hline
\end{tabular}
\end{center}
\end{table}

\begin{figure}[!h]
    \centering
    \subcaptionbox{S}{
        \includegraphics[width=0.3\linewidth]{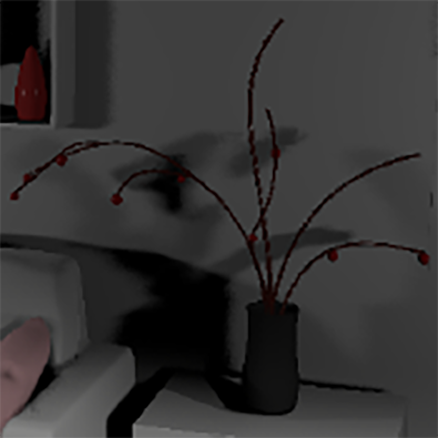}
    }
    \subcaptionbox{M}{
        \includegraphics[width=0.3\linewidth]{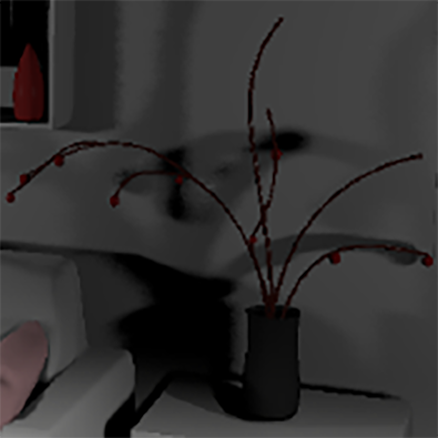}
    }
    \subcaptionbox{L}{
        \includegraphics[width=0.3\linewidth]{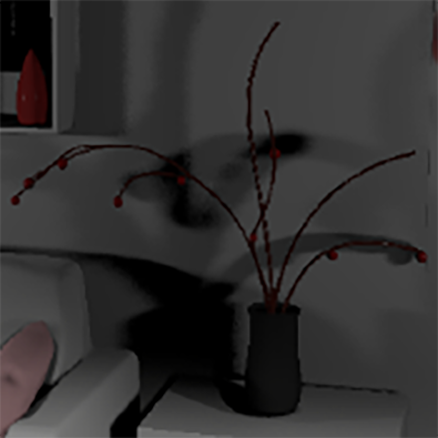}
    }
    \par\smallskip
    \caption{Thin object shadows.}
    \label{fig:resolution}
\end{figure}

However, at $64^3$, the jump flooding is unable to capture some surfaces of thin objects as seen in \autoref{fig:resolution} where the plant's shadow is disjointed and incomplete. Increasing the ray count gives little noticeable improvement as the ray mask is too small. 

\subsubsection{Ray Mask Size}

\begin{figure}[!h]
    \centering
    \subcaptionbox{$d$ = 0.05}{
        \includegraphics[width=0.3\linewidth]{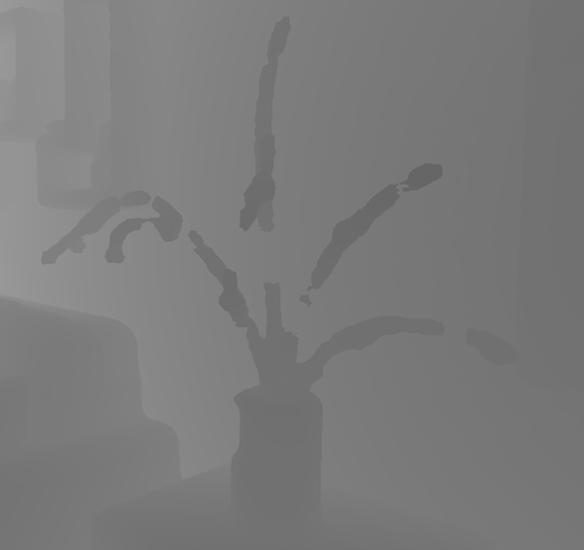}
    }
    \subcaptionbox{$d$ = 0.1}{
        \includegraphics[width=0.3\linewidth]{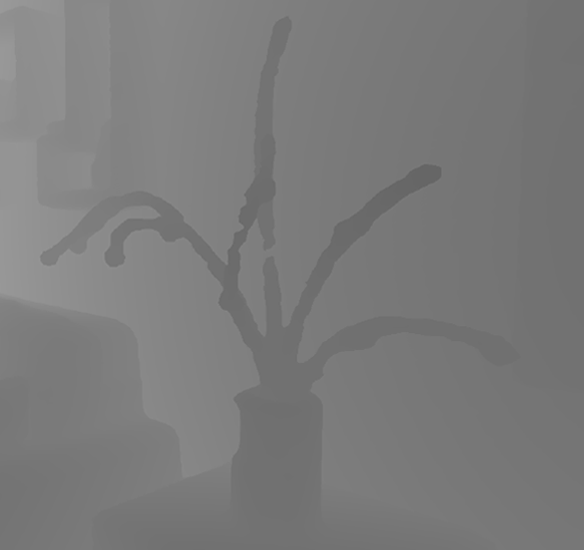}
    }
    \subcaptionbox{$d$ = 0.5}{
        \includegraphics[width=0.3\linewidth]{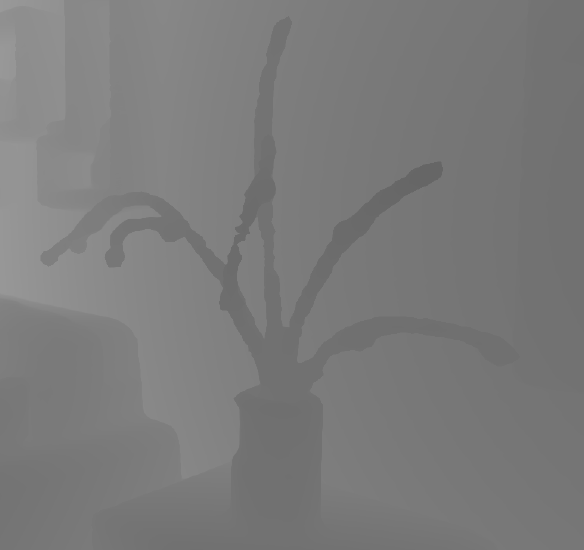}
    }
    \par\smallskip
    \caption{Thin surfaces.}
    \label{fig:ray-mask-size}
\end{figure}

We measure the effect of distance $d$ which determines the size of the ray mask in \autoref{tab:ray-mask-size}. Here, $d$ = $\infty$ corresponds to the case where every texel in the \emph{fine} SDF is ray-traced. As shown in \autoref{fig:ray-mask-size}, there are noticeable holes in the SDF for $d$ = 0.05 as the low resolution of the \emph{coarse} SDF makes it difficult to voxelize thin surfaces. Increasing $d$ to 0.1 fills up the missing holes as we use a more aggressive ray mask which can detect the thin surfaces. However, there is less qualitative difference from $d$ = 0.1 to $d$ = 0.5. Weighing it against the decrease in performance, it appears that $d$ = 0.1 is most suitable for the shot. 

\begin{table}[!h]
\begin{center}
    \caption{Size M SDF GPU timings (ms).}
    \label{tab:ray-mask-size}
\begin{tabular}{|l|c|c|c|c|c|c|}
    \hline
    \diagbox{$x$}{$d$} & 0.01 & 0.05 & 0.1   & 0.5   & $\infty$ \\ \hline
    1                  & 0.8  & 1.28 & 1.63  & 3.12  & 4.08     \\ \hline
    5                  & 1.43 & 3.17 & 4.82  & 11.71 & 16.15    \\ \hline
    10                 & 2.18 & 5.41 & 8.36  & 22.25 & 30.69    \\ \hline
    15                 & 3.17 & 7.7  & 12.42 & 32.7  & 46.0     \\ \hline
\end{tabular}
\end{center}
\end{table}

\subsection{Graphics Quality}

\begin{figure}[!h]
    \centering
    \subcaptionbox{Shadow Map}{
        \includegraphics[width=0.3\linewidth]{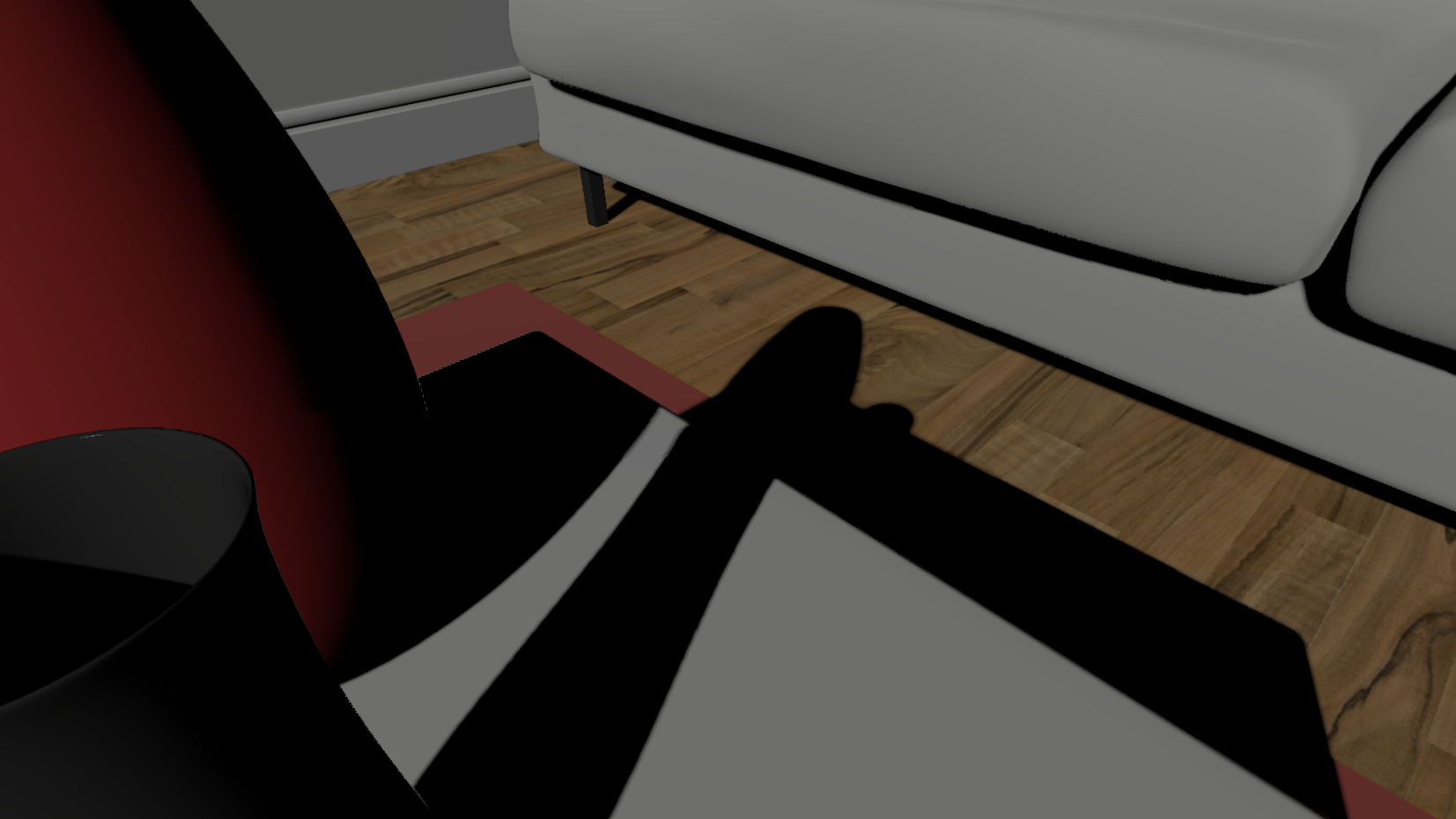}
    }
    \subcaptionbox{RTSDF}{
        \includegraphics[width=0.3\linewidth]{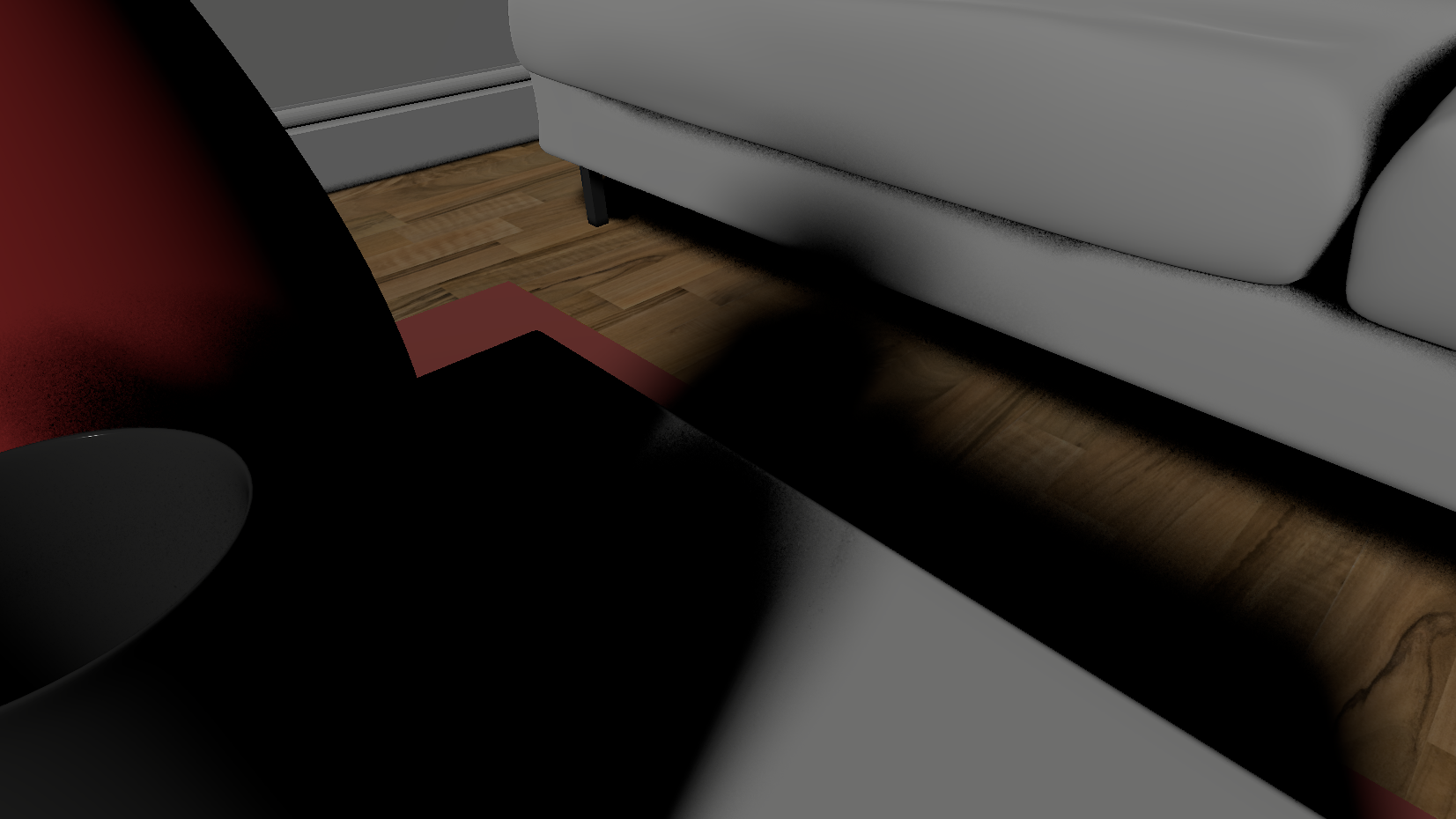}
    }
    \subcaptionbox{Distributed RT}{
        \includegraphics[width=0.3\linewidth]{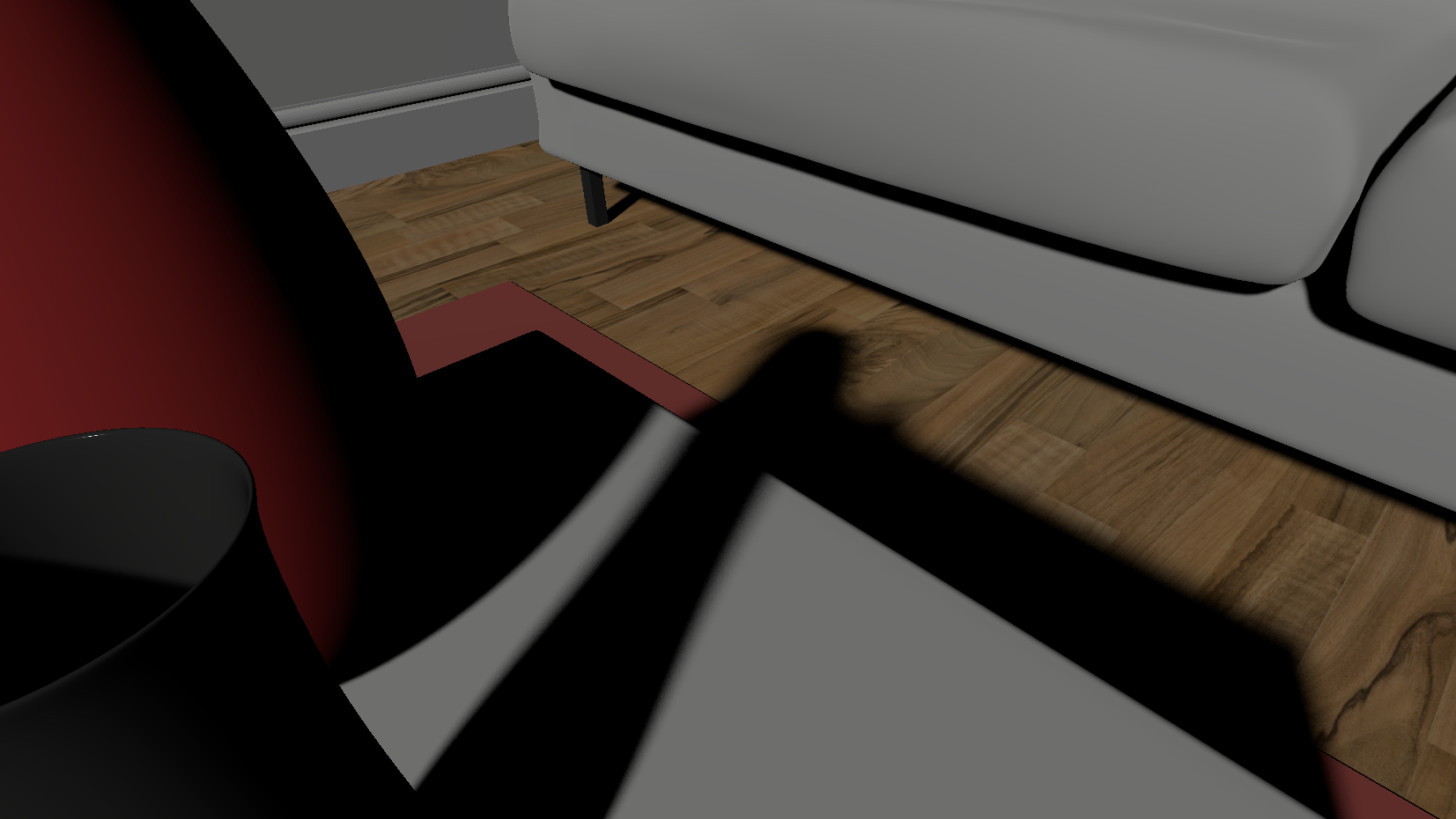}
    }
    \par\smallskip
    \caption{Penumbra.}
    \label{fig:penumbra}
\end{figure}

We compare the smoothness of the soft shadow penumbra generated with our approach against shadow mapping as well as the ground truth as shown in \autoref{fig:penumbra} with one directional light. RTSDF generates a plausible penumbra while the penumbra from shadow mapping is hardly visible. For shadow mapping, a 15px $\times$ 15px kernel was used to generate soft shadows by blurring with EVSM filtering. It could be the case that the shadow map resolution is too low in the foreground for better quality penumbra. We manage to recreate the details of the penumbra more clearly such that it is closer to the distributed ray tracing reference. Our penumbra appears to be larger than the ground truth as a result of adding the small bias in \autoref{sec:holes} to prevent holes in thin surfaces. 

\subsection{Limitations}

\begin{figure}[!h]
    \centering
    \includegraphics[width=\linewidth]{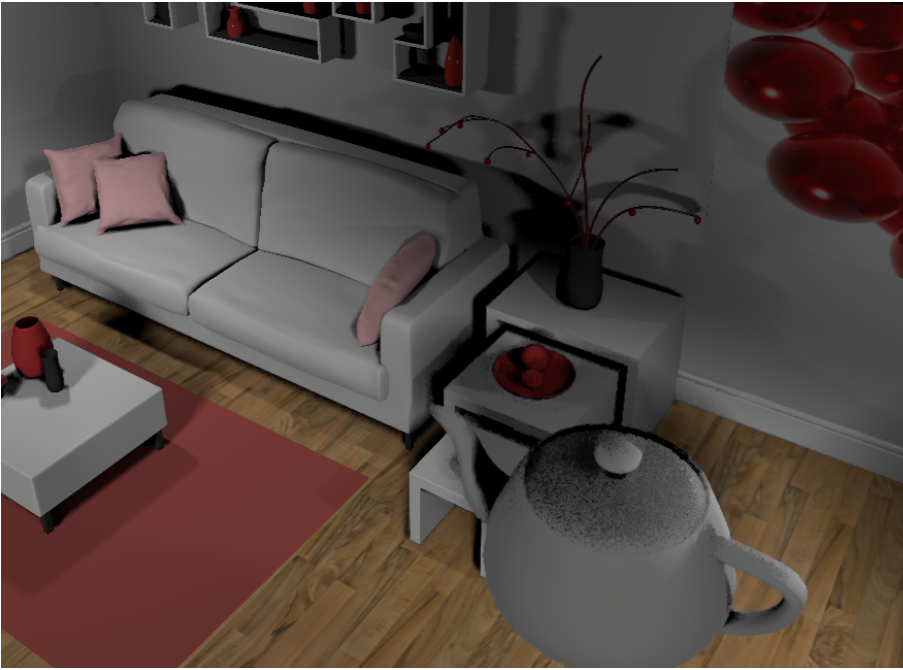}
    \caption{Self intersection of moving teapot.}
    \label{fig:self-intersection}
\end{figure}

Due to a small amount of ghosting of the SDF, our rays shot towards the light source may intersect with the ghosted surface representation of the object. This results in incorrectly occluded areas such as the top of the teapot in \autoref{fig:self-intersection}. For static regions, this artifact would not be present.

\begin{figure}[!h]
     \centering
     \includegraphics[width=\linewidth]{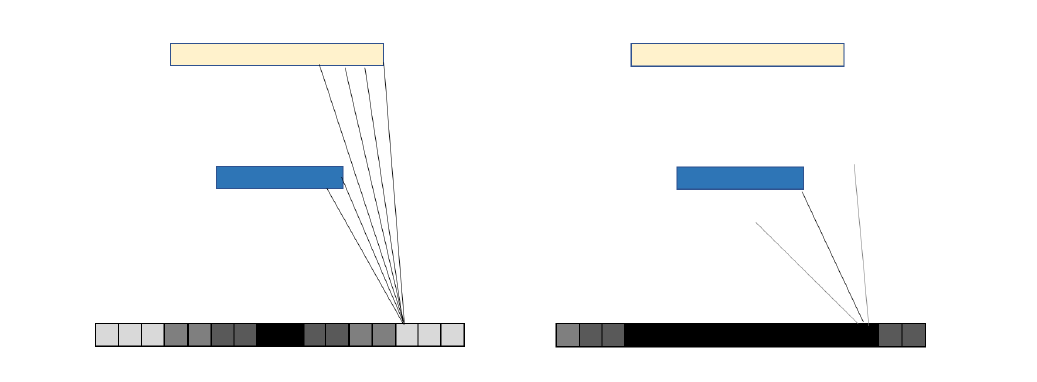}
     \caption{Reference (left) vs penumbra estimation (right).}
     \label{fig:gt-penumbra}
 \end{figure}

Though not a focus of the paper, a key thing to note is also the accuracy of the shadows. Our soft shadow algorithm widens the penumbra and does not accurately calculate umbra as shown in \autoref{fig:gt-penumbra}. On the left, for the ground truth, we shoot multiple visibility rays towards the area light source and hence have an occlusion factor of more than 0. However, for the SDF raymarching estimation, we raymarch a single ray towards the centre of the light source and record an occlusion factor of 0. Consequently, this technique is referred to as penumbra widening shadows by \citet{Aaltonen:2018:AGT}. 

\begin{figure}[!h]
    \centering
    \subcaptionbox{RTSDF}{
        \includegraphics[width=0.47\linewidth]{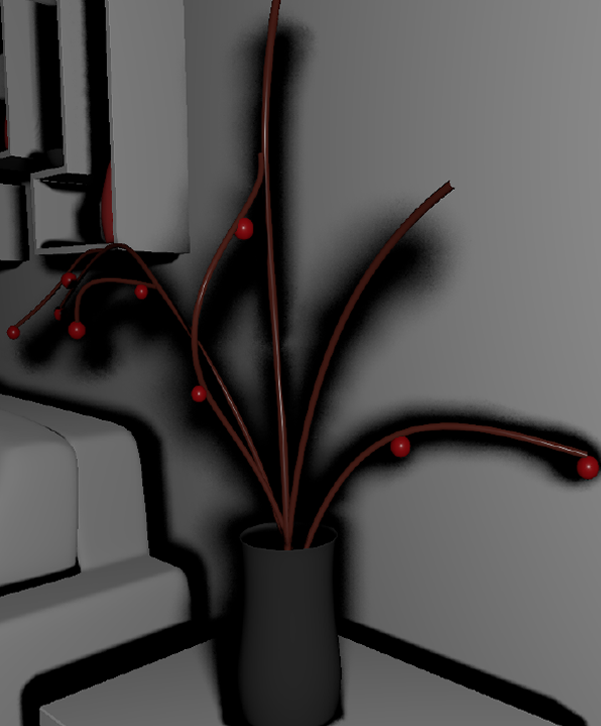}
    }
    \subcaptionbox{Ray Trace (1 per frame)}{
        \includegraphics[width=0.47\linewidth]{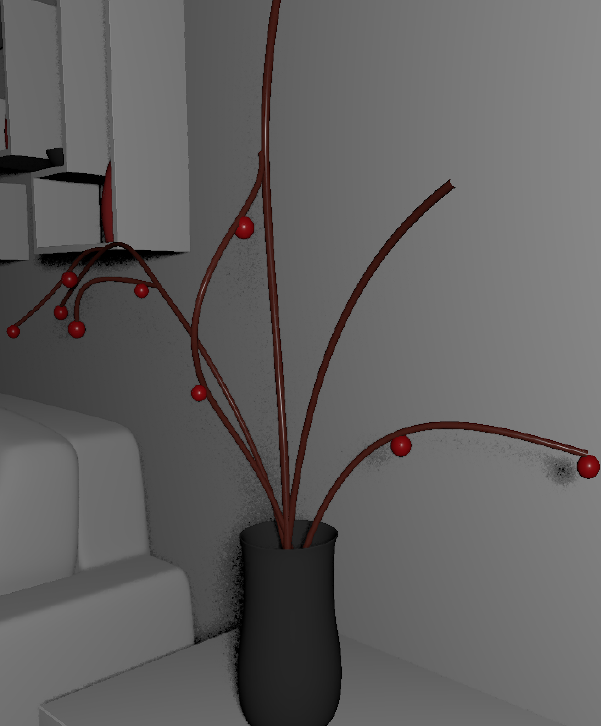}
    }
    \par\smallskip
    \subcaptionbox{Distributed Ray Trace}{
        \includegraphics[width=0.47\linewidth]{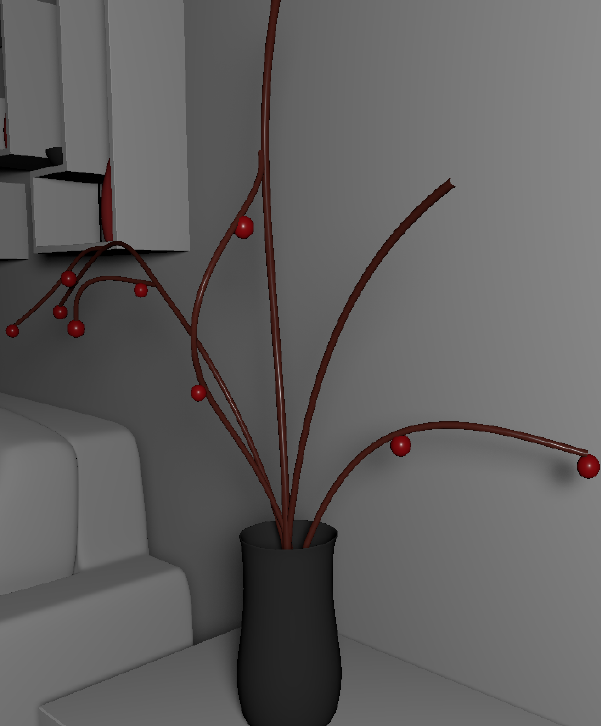}
    }
    \subcaptionbox{SDF-Simulated Umbra}{
        \includegraphics[width=0.47\linewidth]{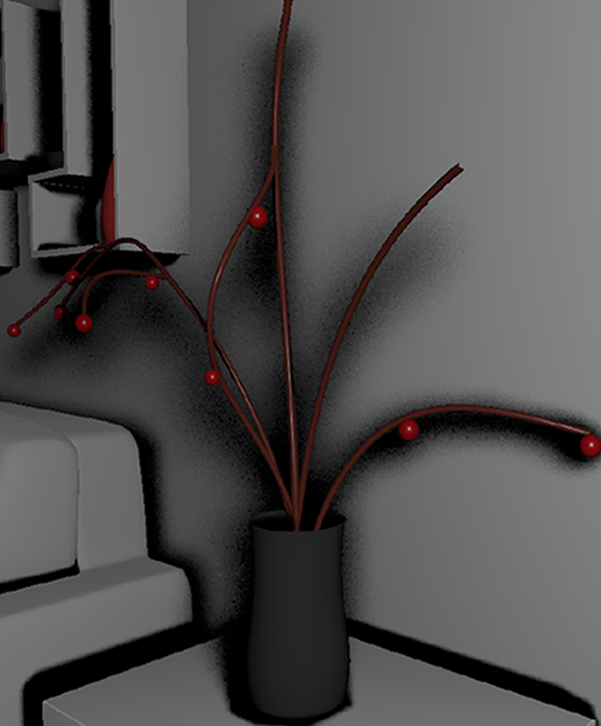}
    }
    \par\smallskip
    \caption{Umbra.}
    \label{fig:simulate-umbra}
\end{figure}

This becomes noticeable as we increase the radius of our area light source, and accompanied by the increased thickness of the surface representation causes our shadows to substantially thicken and darken. More research could be done to investigate potential ways to calculate the shadow umbra given an SDF to achieve less noisy results than a ray trace counterpart. For example, as shown in \autoref{fig:simulate-umbra}, we could perform random sampling towards the light and weight the occlusion factor based on the distance to the closest object along the ray towards the light.

\subsection{Future Work}

\subsubsection{SDF}

Currently, the setup is fine-tuned for a small simple scene. However, it gets more complex when accounting for larger scenes. Having a single SDF for the entire scene is not feasible due to the memory and computation requirements to get a good resolution per $m^3$ 
of the scene. A potential strategy would be to have a clip map explained by \citet{Panteleev:2014:PRV}, with multiple resolutions based on the camera’s position and view direction. In regions outside of the camera’s view frustum, a lower resolution SDF can be calculated which would reduce the cost of rendering. Furthermore, these regions could potentially use only jump flooding instead of adding ray tracing.

As of the date of submission, the acceleration structures generated by the GPU drivers which support the DirectX Raytracing API are not accessible in code. They are bounding volume hierarchies that are used for ray tracing and the API exposed only allows for ray tracing queries. As noted by \citet{Quilez:2019:SBV}, proximity information can be queried from the acceleration structures and potentially be used for rendering effects. These structures could also help aid better reconstruction of the SDF.


\subsubsection{Soft Shadows}

\begin{figure}[!h]
    \centering
    \subcaptionbox{Before TAA}{
        \includegraphics[width=0.47\linewidth]{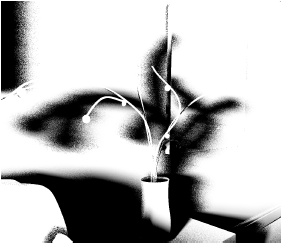}
    }
    \subcaptionbox{After TAA}{
        \includegraphics[width=0.47\linewidth]{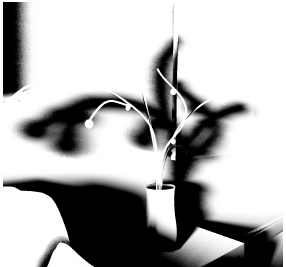}
    }
    \par\smallskip
    \caption{Noise.}
    \label{fig:noise-taa}
\end{figure}

Currently, we apply TAA to reduce noise in the image as shown in \autoref{fig:noise-taa}. However, noise is not eliminated. We could adopt ideas from existing techniques such as spatiotemporal variance-guided filtering \citep{Schied:2017:SVF} which uses screen space blurs and temporal accumulation to remove noise from path tracing samples. With more intelligent filtering of the final result, we could potentially get away with a larger raymarch step introduced to remove banding artifacts and improve performance.

Raymarching is an expensive operation given the required number of samples per pixel. We could also optimize this by combining our soft shadows with a shadow map pass where we decide to raymarch only if not in shadow.

While not explored, we note that since we have an SDF approximation of the scene, there is the potential of representing translucent surfaces to generate translucent soft shadows. The main initial difficulty would be identifying the surface intersected when raymarching a distance field. We could potentially store surface information in the distance field and perform optimization by using lookup tables.
\section{\uppercase{Conclusion}}

We developed a novel technique that combines jump flooding and ray tracing to generate SDFs in real-time with plausible results for soft shadowing and exposed values that trade-off between performance and quality of the SDF generated which would be useful when targeting hardware of different specifications. Our approach can handle scenes with dynamic objects and produce penumbra that is smoother than shadow mapping but cleaner than distributed ray tracing. 

\section*{\uppercase{Acknowledgements}}
We thank \citet{Wyman:2018:IDR} for the Falcor scene file of \textsc{The Modern Living Room} (\href{https://creativecommons.org/licenses/by/3.0/}{CC BY}). This work is supported by the Singapore Ministry of Education Academic Research grant T1 251RES1812, “Dynamic Hybrid Real-time Rendering with Hardware Accelerated Ray-tracing and Rasterization for Interactive Applications”. 

\bibliographystyle{apalike}
{\small
\bibliography{rtsdf}}

\end{document}